\documentclass[10pt,journal,twoside,final]{IEEEtran}

\IEEEoverridecommandlockouts

\makeatletter
\newcommand{\biggg}[1]{{\hbox{$\left#1\vbox to 20.5pt{}\right.\n@space$}}}
\newcommand{\Biggg}[1]{{\hbox{$\left#1\vbox to 23.5pt{}\right.\n@space$}}}
\newcommand{\bigggg}[1]{{\hbox{$\left#1\vbox to 26.5pt{}\right.\n@space$}}}
\newcommand{\Bigggg}[1]{{\hbox{$\left#1\vbox to 29.5pt{}\right.\n@space$}}}
\newcommand{\biggggg}[1]{{\hbox{$\left#1\vbox to 32.5pt{}\right.\n@space$}}}
\newcommand{\Biggggg}[1]{{\hbox{$\left#1\vbox to 35.5pt{}\right.\n@space$}}}
\newcommand{\bigggggg}[1]{{\hbox{$\left#1\vbox to 38.5pt{}\right.\n@space$}}}
\newcommand{\Bigggggg}[1]{{\hbox{$\left#1\vbox to 41.5pt{}\right.\n@space$}}}
\makeatother

\usepackage{bm,cite,algorithm,algorithmic,float,amsmath,amssymb}
\usepackage[dvips]{graphicx}
\usepackage{subfigure,amsfonts}
\usepackage{mdwmath}
\usepackage{mdwtab}
\usepackage{xcolor,color}
\usepackage{cool}
\usepackage{balance}
\usepackage{diagbox}
\floatname{algorithm}{Algorithm}

\makeatletter
\renewcommand\paragraph{\@startsection{paragraph}{4}{\z@}%
            {-2.5ex\@plus -1ex \@minus -.25ex}%
            {1.25ex \@plus .25ex}%
            {\normalfont\normalsize\itshape}}
\makeatother
\usepackage{epstopdf}

\usepackage[style=0]{mdframed}
\usepackage{showexpl}

\mdfdefinestyle{exampledefault}{%
rightline=true,innerleftmargin=10,innerrightmargin=10,
frametitlerule=true,frametitlerulecolor=green,
frametitlebackgroundcolor=yellow,
frametitlerulewidth=2pt}

\allowdisplaybreaks

\begin{document}

\title{\huge{Covert Communications in Active-IOS Aided Uplink\\ 
NOMA Systems With Full-Duplex Receiver}}

\author{
Xueyu Kang, Nan Qi,  
Lu Lv,  
Alexandros-Apostolos A. Boulogeorgos,\\
 
Theodoros A. Tsiftsis,  
and Hongwu~Liu

\thanks{X. Kang and H. Liu are with the School of Information Science and Electrical Engineering, Shandong Jiaotong University, Jinan 250357, China (e-mail: 1113198984@qq.com, liuhongwu@sdjtu.edu.cn).}
\thanks{N. Qi is with the Key Laboratory of Dynamic Cognitive System of Electromagnetic Spectrum Space, Nanjing University of Aeronautics and Astronautics, Nanjing 210016, China, and also with the State Key Laboratory of Integrated Services Networks, Xidian University, Xi’an 710071, China (e-mail: nanqi.commun@gmail.com).}
\thanks{L. Lv is with the School of Telecommunications Engineering, Xidian University, Xi'an 710071, China (e-mail: lulv@xidian.edu.cn).}
\thanks{A.-A. A. Boulogeorgos is with the Department of Electrical and Computer Engineering, University of Western Macedonia, 50100 Kozani, Greece (e-mail: al.boulogeorgos@ieee.org).}
\thanks{T. A. Tsiftsis is with the Department of Informatics and Telecommunications, University of Thessaly, Lamia 35100, Greece, and also with the Department of Electrical and Electronic Engineering, University of Nottingham Ningbo China, Ningbo 315100, China (e-mail: tsiftsis@uth.gr).} 
}

\maketitle
\setcounter{page}{1} 
\begin{abstract} 
In this paper, an active intelligent omni-surface (A-IOS) is deployed to aid uplink transmissions in a non-orthogonal multiple access (NOMA) system. In order to shelter the covert signal embedded in the superposition transmissions, a multi-antenna full-duplex (FD) receiver is utilized at the base-station to recover signal in addition to jamming the warden. With the aim of maximizing the covert rate, the FD transmit and receive beamforming, A-IOS refraction and reflection beamforming, NOMA transmit power, and FD jamming power are jointly optimized. To tackle the non-convex covert rate maximization problem subject to the highly coupled system parameters, an alternating optimization algorithm is designed to iteratively solve the decoupled sub-problems of optimizing the system parameters. The optimal solutions for the sub-problems of the NOMA transmit power and FD jamming power optimizations are derived in closed-form. To tackle the rank-one constrained non-convex fractional programming of the A-IOS beamforming and FD beamforming, a penalized Dinkelbach transformation approach is proposed to resort to the optimal solutions via semidefinite programming. 
Numerical results clarify that the deployment of the A-IOS significantly improves the covert rate compared with the passive-IOS aided uplink NOMA system. It is also found that the proposed scheme provides better covert communication performance with the optimized NOMA transmit power and FD jamming power compared with the benchmark schemes.  
\end{abstract}

\section{Introduction}

As a promising technique to provide full-space wireless coverage, intelligent omni-surface (IOS) has significantly expanded wireless communication ranges by refracting and reflecting the incident signals to the users located on both sides \cite{RIS_6G,STAR_360}. 
Technically, an IOS is a double-sided planar metasurface composed of a large number of low-cost refraction and reflection elements, where the beamforming on both sides of the IOS can be independently adjusted to achieve different design goals, such as creating additional propagation paths, strengthening and weakening signal powers on desired directions \cite{BIOS}.  To overcome the ``multiplicative-fading'' effect, active load impedances were introduced in active-IOS (A-IOS) for signal amplification on the refraction and reflection sides, respectively, with the improved coverage performance for wireless communication systems \cite{AIOS_Vehicular, EE_STAR_RIS_C, AIOS_Beamforming}. Since the incident signals at the IOS are naturally superimposed in the power-domain, the IOS is preferable to be deployed in non-orthogonal multiple access (NOMA) systems to improve spectral efficiency and network connectivity \cite{AIOS_Beamforming, STAR_RIS_NOMA_Coverage}.   

Although leveraging IOS and NOMA is a win-win strategy, adversaries may achieve similar performance gains as the legitimate users in IOS aided NOMA systems and the corresponding security and privacy issues have attracted significant research interest in both civilian and military applications \cite{STAR_RIS_NOMA_Covert, STAR_RIS_NOMA_Covert_XiaoHan, STAR_RIS_IoT_Covert}. As a new security paradigm to provide a higher level of security than physical layer security, covert communications were investigated to conceal the communications behaviors in IOS aided NOMA systems \cite{STAR_RIS_NOMA_Covert, STAR_RIS_NOMA_Covert_XiaoHan, STAR_RIS_IoT_Covert}. Besides optimizing the refraction and reflection beamforming at the IOS, full-duplex (FD) receivers were further introduced to transmit artificial noise (AN) signal for improving the covert communication performance 
\cite{STAR_RIS_FD_Receiver, STAR_RIS_FD_RSMA}. However, most of the existing works considered covert communications in IOS aided downlink NOMA systems \cite{STAR_RIS_NOMA_Covert, STAR_RIS_NOMA_Covert_XiaoHan, STAR_RIS_IoT_Covert, STAR_RIS_FD_Receiver}, where the small-sized receivers limited the numbers of the FD transmit and receive antennas, such that the covert communication performance enhancements were limited. On the contrary, in uplink NOMA systems, receivers a.k.a. base-stations (BSs), in general have enough space to deploy the FD transmit and receive antennas, which enables a new feasibility to improve the covert communication performance for IOS aided uplink NOMA systems. To the best of our knowledge, covert communications in IOS aided uplink NOMA systems are still in an early stage. How to employ IOS, especially employing A-IOS and FD receiver, to enhance the covert communication performance in uplink NOMA systems is still unknown.  

In this work, we propose a covert communication scheme for an IOS aided uplink NOMA system. Specifically, we employ the A-IOS to enhance the covert communication performance by optimizing the refraction and reflection beamforming at the A-IOS and the transmit power allocation at the NOMA users.  To further enhance the covert communication performance, the FD capability is applied at the BS to transmit AN signal toward the warden. Constrained by the power budgets at the FD BS, A-IOS, and NOMA users, we formulate the covert rate maximization problem by jointly optimizing the NOMA transmit power allocation, jamming power allocation, FD beamforming, and A-IOS beamforming subject to the covertness and quality of service (QoS) constraints. The main contributions of this paper include: 
1) We first decouple the original non-convex covert rate maximization problem into several sub-problems
and derive the expressions for the optimal solutions for the NOMA transmit power and FD jamming power optimization sub-problems. Then, we tackle the original problem by solving the decoupled sub-problems using the alternating optimization (AO) technique.  2) For A-IOS beamforming and FD beamforming sub-problems, a penalized Dinkeltach transformation is proposed to tackle the rank-one constrained non-convex fractional programming and obtain the corresponding optimal solutions via semidefinite programming. 3) Verified by the simulation results, the proposed AO algorithm significantly enhances the  covert communication performance by employing not only the A-IOS, but also the FD receiver with deliberate jamming.   
   
\section{System Model}

We consider an A-IOS aided uplink NOMA system consisting of a covert user (Alice), a public user (Grace), a warden (Willie), a FD receiver (Bob), and an A-IOS consisting of $K$ elements. We assume that Alice, Grace, and Willie are equipped with a single antenna, respectively, Bob is equipped with $M$ transmit antennas and $M$ receive antennas, and the A-IOS works in the energy splitting (ES) mode \cite{STAR_360}, i.e, $K$ elements work simultaneously in the refraction and reflection modes such that the incident signal on each element is split into the refraction and reflection parts, respectively. Furthermore, Alice, who has the covertness requirements to hide its communications to Willie, is located on the refraction side of the A-IOS, while Grace, who only needs to satisfy its QoS requirements, is located on the reflection side of the A-IOS. We assume that the Alice-Bob and Grace-Bob direct-links are unavailable due to severe blockages.  
Let ${\textbf{H}}_{ob} \in \mathbb{C}^{M \times K}$ and ${\textbf{H}}_{bo} \in \mathbb{C}^{K \times M}$ denote the channel coefficients from the A-IOS to Bob's receive antennas and from Bob's transmit antennas to the A-IOS, respectively. 
Let ${\textbf{h}}_{ao} \in \mathbb{C}^{K \times 1}$, ${\textbf{h}}_{go} \in \mathbb{C}^{K \times 1}$, ${\textbf{h}}_{ow} \in \mathbb{C}^{1 \times K}$, and ${\textbf{h}}_{bw} \in \mathbb{C}^{1 \times M}$ denote the channel coefficients of the Alice-A-IOS, Grace-A-IOS, A-IOS-Willie, and Bob-Willie links, respectively. 
Similar to works in \cite{BIOS,  STAR_RIS_NOMA_Covert_XiaoHan, STAR_RIS_IoT_Covert}, we assume that all channels experience quasi-static block 
fading, Willie has perfect channel state information (CSI) of the A-IOS related links, Bob has instantaneous CSI of the Alice-A-IOS-Bob and Grace-A-IOS-Bob links, and statistical CSI of the channels associated with Willie.  

\subsection{Uplink NOMA Transmissions}

In a transmission block, the information signals transmitted by Alice and Grace are denoted by $s_a$ and $s_g$, respectively, while the AN signal transmitted by Bob for the jamming purpose is denoted by $s_j$. We assume that the signals satisfy $\mathbb{E} \left(|s_a|^2 \right) = \mathbb{E} \left(|s_g|^2 \right) = \mathbb{E} \left(|s_j|^2 \right) = 1$. Then, the incident signal at the A-IOS can be written as 
$ {\textbf{y}}_o = \textbf{h}_{ao} \sqrt{P_a} s_a + {\textbf{h}}_{go} \sqrt{P_g} s_{g} + \textbf{H}_{bo}\textbf{w}_t\sqrt{P_j}s_j $,
where $P_a$, $P_g$, and $P_j$ are the transmit powers of Alice, Grace, and Bob, respectively, and $\textbf{w}_t \in \mathbb{C}^{M \times 1}$ is the transmit beamforming vector at Bob. At the A-IOS, the incident signal will be refracted and reflected, simultaneously, to Bob, while Bob transmits the jamming signal $s_j$ at the same time, which result in the received signal at Bob as follows:  
\begin{align}
y_{b} = ~& \textbf{w}_{r}^{H} {\textbf{H}}_{ob} {\bf{\Theta}}_t {\textbf{h}}_{ao}  \sqrt{P_a}s_{a} + \textbf{w}_{r}^{H}  {\textbf{H}}_{ob} {\bf{\Theta}}_r {\textbf{h}}_{go} \sqrt{P_g}s_g  \nonumber \\& +  \textbf{w}_{r}^{H}(\textbf{H}_{bb} + \textbf{H}_{ob}{\bf{\Theta}}_r\textbf{H}_{bo} ) \textbf{w}_t \sqrt{P_j}s_j \nonumber \\ & + 
\textbf{w}_{r}^{H} \big({\textbf{H}}_{ob} {\bf{\Theta}}_t {\textbf{z}}_o  + {\textbf{H}}_{ob} {\bf{\Theta}}_r {\textbf{z}}_o +{\textbf{z}}_{b} \big),
\end{align}
where $\textbf{w}_r \in \mathbb{C}^{M \times 1}$ is the receive beamforming vector at Bob, $\textbf{H}_{bb} \sim \mathcal{CN} (0, \textbf{I}_M)$ is the self-interference (SI) channel at Bob, 
$\textbf{z}_o \sim \mathcal{CN} (0, \sigma_{o}^{2} {\textbf{I}}_{K})$ and ${\textbf{z}}_{b} \sim \mathcal{CN} (0, \sigma_{b}^{2} {\textbf{I}}_M)$ are the additive noises at the A-IOS and Bob, respectively, and ${\bf{\Theta}}_i = {\rm{diag}} \big(  \alpha_{1}^{i} e^{j\theta_{1}^{i}},\cdots,  \alpha_{K}^{i} e^{j\theta_{K}^{i}} \big), i \in \{t,r \}$ denotes the A-IOS’s refraction and reflection coefficients with  $0 \le \alpha_{k}^{t}, \alpha_{k}^{r} \le a_{k}^{\max}$, $k \in \{1, 2, \cdots, K\}$, representing the refraction and reflection amplitudes of the $k$th  element and $a_{k}^{\max}$ is the maximum processing amplitude \cite{AIOS_Beamforming}.
On the other hand, the power of the departure signal from the A-IOS is limited by
\begin{align}
    ~&  P_a \|{\bf{\Theta}}_t {\textbf{h}}_{ao}\|^2 + P_g \|{\bf{\Theta}}_r {\textbf{h}}_{go}\|^2 + P_j \|{\bf{\Theta}}_r \textbf{H}_{bo} \textbf{w}_t \|^2   \nonumber 
    \\ ~& + \|{\bf{\Theta}}_t\|_{F}^2 \sigma_{o}^{2} + \|{\bf{\Theta}}_r\|_{F}^2 \sigma_{o}^{2}  \le P_{o}^{\max},  \label{P_s}
\end{align}
where $P_{o}^{\max}$ is the power budget at the A-IOS.

In uplink NOMA systems, the receiver generally decodes the users with better channel conditions. In order to maximize Alice's covert rate, the composite channels are arranged as $\|\textbf{H}_{ob} {\bf{\Theta}}_r \textbf{h}_{go}\|^2 \ge \|\textbf{H}_{ob} {\bf{\Theta}}_t \textbf{h}_{ao}\|^2$, such that $s_a$ is decoded at the last stage of the successive interference cancellation (SIC) processing without severe inter-user interference. 
Thus, the achievable rates of Alice and Grace can be expressed as
$R_a = \log_2 (1+ \gamma_a)$ and $R_g = \log_2 ( 1+ \gamma_g )$,
respectively, where $\gamma_a = \tfrac{ P_a |\textbf{g}_{ab}|^2}{{\bf{\Omega}} + \sigma_{b}^{2}}$ and 
$\gamma_g = \tfrac{P_g|\textbf{g}_{gb}|^2}{P_a|\textbf{g}_{ab}|^2 + {\bf{\Omega}} +  \sigma_{b}^{2}}$ are the received signal-to-interference-plus-noise ratios (SINRs) with $|\textbf{g}_{ab}|^2 = |\textbf{w}_{r}^{H} \textbf{H}_{ob} {\bf{\Theta}}_t \textbf{h}_{ao}|^2$, $|\textbf{g}_{gb}|^2 = |\textbf{w}_{r}^{H} \textbf{H}_{ob} {\bf{\Theta}}_r \textbf{h}_{go}|^2$,  
${\bf{\Omega}} = \|{\textbf{w}}_{r}^{H} {\textbf{H}}_{ob} {\bf{\Theta}}_t\|^2 \sigma_{o}^{2} + \|{\textbf{w}}_{r}^{H} {\textbf{H}}_{ob} {\bf{\Theta}}_r\|^2 \sigma_{o}^{2} +  \phi P_j |\textbf{w}_{r}^{H} (\textbf{H}_{bb} + \textbf{H}_{ob}{\bf{\Theta}}_r \textbf{H}_{bo} ) \textbf{w}_t|^2$, and $\phi \in [0,1]$ is the SI cancellation level at Bob \cite{STAR_RIS_FD_RSMA}. 

\subsection{Willie’s Detection}

Let $\mathcal{H}_0$ and $\mathcal{H}_1$ denote the two hypotheses representing that Alice is not transmitting and transmitting, respectively, under which the received signal at Willie can be expressed as:  
\begin{align} 
\mathcal{H}_0 : {\textbf{y}}_{w} = ~&{\textbf{h}}_{ow} {\bf{\Theta}}_r {\textbf{h}}_{go} \sqrt{P_g}s_g +  {\textbf{h}}_{ow} {\bf{\Theta}}_r {\textbf{z}}_o \nonumber 
\\ 
& + ({\textbf{h}}_{ow} {\bf{\Theta}}_r {\textbf{H}}_{bo} + \textbf{h}_{bw}) \textbf{w}_t \sqrt{P_j}s_j + z_{w}, \label{W:H0}
\end{align}
\begin{align} 
\mathcal{H}_1: {\textbf{y}}_{w}  = ~&  {\textbf{h}}_{ow} {\bf{\Theta}}_t {\textbf{h}}_{ao} \sqrt{P_a}s_a  +  {\textbf{h}}_{ow} {\bf{\Theta}}_r {\textbf{h}}_{go} \sqrt{P_g}s_g  \nonumber \\  
 ~& + ({\textbf{h}}_{ow} {\bf{\Theta}}_r {\textbf{H}}_{bo} + \textbf{h}_{bw}) \textbf{w}_t \sqrt{P_j}s_j + {\textbf{h}}_{ow} {\bf{\Theta}}_t {\textbf{z}}_o  \nonumber \\
 ~& + {\textbf{h}}_{ow} {\bf{\Theta}}_r {\textbf{z}}_o + z_{w} , \label{W:H1}
\end{align}
where $z_w \sim \mathcal{CN} (0, \sigma_{w}^{2})$ is the additive noise at Willie.
To distinguish between the above two hypotheses,  Willie adopts the Neyman-Person criterion and obtains its optimal likelihood ratio test as    
${\small |{\textbf{y}}_w|^2 \mathop{\gtrless}\limits_{\mathcal{D}_0 }^{\mathcal{D}_1} \varphi^*}$ \cite{Zhumiaomiao},
where $\mathcal{D}_1$ and $\mathcal{D}_0$ indicate whether Alice is transmitting the signal to Bob or not, respectively, and 
$\varphi^* = \frac{\lambda_1 \lambda_0}{\lambda_1 - \lambda_0} \ln \frac{\lambda_1}{\lambda_0} > 0$ is the optimal detection threshold with 
$\lambda_{0} = P_g |{\textbf{g}}_{gw}|^2 + P_j |\textbf{g}_{bw} \textbf{w}_t|^2 + \|{\textbf{h}}_{ow} {\bf{\Theta}}_r\|^2 \sigma_{o}^{2} + \sigma_{w}^{2}$,  $\lambda_{1} = P_g |{\textbf{g}}_{gw}|^2 + P_j |\textbf{g}_{bw} \textbf{w}_t |^2  + P_a|{\textbf{g}}_{aw}|^2 + \|{\textbf{h}}_{ow} {\bf{\Theta}}_t\|^2 \sigma_{o}^{2} +\|{\textbf{h}}_{ow} {\bf{\Theta}}_r\|^2 \sigma_{o}^{2} +\sigma_{w}^{2}$, ${\textbf{g}}_{gw} = {\textbf{h}}_{ow} {\bf{\Theta}}_r {\textbf{h}}_{go}$, ${\textbf{g}}_{aw} = {\textbf{h}}_{ow} {\bf{\Theta}}_t {\textbf{h}}_{ao}$, and ${\textbf{g}}_{bw} = {\textbf{h}}_{ow} {\bf{\Theta}}_r {\textbf{H}}_{bo} + \textbf{h}_{bw}$.
Let ${\rm{Pr}}(\mathcal{D}_1 | \mathcal{H}_0)$ and ${\rm{Pr}}(\mathcal{D}_0 | \mathcal{H}_1)$ denote Willie's false alarm and miss detection probabilities, respectively, Willie's minimum detection error probability (MDEP) can be derived as \cite{Zhumiaomiao}: 
\begin{align}
  \xi_{w}^{*} =~& {\rm{Pr}}(\mathcal{D}_1 | \mathcal{H}_0) + {\rm{Pr}}(\mathcal{D}_0 | \mathcal{H}_1) \nonumber \\ =~& 1 + ( \lambda_1 / \lambda_0)^{-\frac{\lambda_1}{\lambda_1 - \lambda_0}} - ( \lambda_1 / \lambda_0)^{-\frac{\lambda_0}{\lambda_1 - \lambda_0}}. \label{xi} 
\end{align}
Since further processing of $\xi_{w}^{*}$ involving multiple beamforming vectors is extremely difficult, we introduce a lower bound for $\xi_{w}^{*}$ as follows:
\begin{eqnarray}
    \xi_{w}^{*} \ge 1 - \sqrt{\tfrac{1}{2} \mathcal{D} (p_0 ({\textbf{y}}_w)\|p_1 ({\textbf{y}}_w) ) } , \label{xi_LD}
\end{eqnarray}
where the Kullback-Leibler (KL) divergence $\mathcal{D}(p_0 ({\textbf{y}}_w)\| $ $p_1 ({\textbf{y}}_w) )$ is given by
\begin{eqnarray}
\mathcal{D} (p_0 ({\textbf{y}}_w)\|p_1 ({\textbf{y}}_w) ) = \ln (\lambda_1 / \lambda_0) + (\lambda_0 / \lambda_1) - 1 ,  \label{DL}
\end{eqnarray} 
with $p_0 ({\textbf{y}}_w)$ and $p_1 ({\textbf{y}}_w)$ denoting the likelihood functions for Willie's received signals under $\mathcal{H}_0$ and $\mathcal{H}_1$, respectively. To ensure the minimum covertness level $\varepsilon > 0$ required by Alice, Willie's MDEP should satisfy $\xi_{w}^{*} \ge 1 - \varepsilon$. With the aid of \eqref{xi_LD}, $\xi_{w}^{*} \ge 1 - \varepsilon$ can be further derived as $\mathcal{D} (p_0 ({\textbf{y}}_w)\|p_1 ({\textbf{y}}_w) ) \le 2 \varepsilon^2$ and we adopt it as the covertness constraint in the covert communication design. Since the function $f(\lambda) = \ln \lambda +\frac{1}{\lambda} -1$ in \eqref{DL} is monotonically increasing with respect to $\lambda \in [1, +\infty)$, the covertness constraint can be rewritten as:
\begin{align}
    P_a |{\textbf{g}}_{aw}|^2 + \|{\textbf{h}}_{ow} {\bf{\Theta}}_t\|^2 \sigma_{o}^{2} \le (\kappa - 1)~~~~~~~~~~~~~~~~~~ \nonumber \\ \times \big( P_g  |{\textbf{g}}_{gw}|^2 + P_j |\textbf{g}_{bw} \textbf{w}_t|^2 + \|{\textbf{h}}_{ow} {\bf{\Theta}}_r\|^2 \sigma_{o}^{2}+ \sigma_{w}^{2} \big), \label{PbPg_k}
\end{align}
where $\kappa \in [1,+\infty)$ is the root of the equation $f(\lambda) = 2\varepsilon^2$.

\section{Covert Rate Maximization}
 
In this section, a covert rate maximization problem is formulated by jointly designing the NOMA transmit power, FD jamming power, A-IOS refraction and reflection beamforming, and FD transmit and receive beamforming, which is given by
\begin{subequations}
\begin{align}
({\rm{P1}}):~&\mathop{{\max}}\limits_{P_a, P_g, P_j, {\bf{\Theta}}_t, {\bf{\Theta}}_r, {\textbf{w}_r}, {\textbf{w}_t} }R_a \label{P1a}
\\
{\rm{s.t.}}~& P_a \le P_{a}^{{\max}}, P_g \le P_{g}^{{\max}}, \text{~and~} P_j \le P_{j}^{{\max}},\label{P1b}\\
~& \|\textbf{H}_{ob} {\bf{\Theta}}_{r} \textbf{h}_{go} \|^2 \ge \|\textbf{H}_{ob} {\bf{\Theta}}_{t} \textbf{h}_{ao} \|^2, \label{P1c}
\\ ~&
R_g \ge \hat{R}_g, \label{P1d}
\\~& 
\theta_{k}^{r}, \theta_{k}^{t} \in [0,2\pi ), ~\forall k,  \label{P1e}
\\ ~& {\max} \{\alpha_{k}^{t}, \alpha_{k}^{r} \} \le \alpha_{k}^{{\max}}, ~\forall_k, \label{P1f}
\\ ~& \|{\textbf{w}_r}\|^2 = \|{\textbf{w}_t}\|^2 = 1,  \label{P1g} \\ ~& 
\big( (\alpha_{k}^{t})^2 + (\alpha_{k}^{r})^2 \big)(p_{k}^{{\rm{in}}} + \sigma_{o}^{2}) \le p_k, ~\forall_k, \label{P1h}
\\ ~& \sum\nolimits_{k = 1}^{K} p_{k}^{{\rm{in}}} \le P_a\|\textbf{h}_{ao} \|^2 + P_g\|\textbf{h}_{go} \|^2 + P_j\|\textbf{H}_{bo}\textbf{w}_t \|^2,   \label{P1i} \\ ~& \sum\nolimits_{k = 1}^{K} p_k \le P_{{o}}^{{\max}}, \label{P1j} \\ ~&  ~\eqref{P_s}  {\text{~and~}}
\eqref{PbPg_k}. \label{P1k}
\end{align}
\end{subequations}
In problem (P1), constraint \eqref{P1b} is the transmit power budgets at all the nodes with $P_a^{\max}$, $P_g^{\max}$, and $P_{j}^{{\max}}$ denoting the maximum transmit power of Alice, Grace, and Bob, respectively. Constraint \eqref{P1c} guarantees a successful SIC decoding. Constraint \eqref{P1d}  ensures that Grace's QoS requirements with $\hat R_g$ denoting the minimum target rate of Grace. Constraints \eqref{P1e} and \eqref{P1f} limit the A-IOS's phase-shifts and amplitudes, respectively. Constraint \eqref{P1g} limits the powers of the FD beamforming vectors at Bob. Constraint \eqref{P1h} is the power consumption constraint for each element of the A-IOS with $p_{k}^{\rm{in}}$ and $p_{k}$ denoting the incident signal energy and power budget of the $k$th element of the A-IOS, respectively \cite{AIOS_Beamforming}. 
Constraints \eqref{P1i} and \eqref{P1j} limit the sum power of the incident signal and the total power of A-IOS's elements,  respectively \cite{AIOS_Beamforming}. Constraint \eqref{P_s} limits the refraction and reflection total power of the A-IOS and constraint \eqref{PbPg_k} is the covertness requirement.
Since the system parameters in the objective function and constraints of problem (P1) are coupled in a complex manner, it is extremely challenging to solve problem (P1) directly. Therefore, we first decouple problem (P1) into several sub-problems that can be tackled. Then, we propose an AO algorithm to obtain the optimized solution for problem (P1).

For any given ${\bf{\Theta}}_r,{\bf{\Theta}}_t$, ${\textbf{w}_r}$, and ${\textbf{w}_t}$, problem (P1) can be rewritten as:   
\begin{subequations}
\begin{align}
({\rm{P2}}):~&\mathop{{\max}}\limits_{P_a, P_g, P_j}R_a \label{P2a}\\
{\rm{s.t.}}~&\eqref{P1b}, ~\eqref{P1d}, ~\eqref{P_s},  {\text{~and~}}
\eqref{PbPg_k}. \label{P2c}
\end{align}
\end{subequations}

$\!\!\!\!$With any given $P_g$ and $P_j$, 
constraints \eqref{P1b}, \eqref{P1d}, \eqref{P_s}, and \eqref{PbPg_k} can be equivalently written as $P_a \le P_{a}^{{\max}}$, $P_a \le \Xi_1$, $P_a \le \Xi_2$, and $P_a \le \Xi_3$, respectively, where $\Xi_1 = \frac{P_g | {\textbf{g}}_{gb}|^2 / \mu_g - \Omega - \sigma_{b}^{2}}{|{\textbf{g}}_{ab}|^2}$, $\mu_g = 2^{\hat{R}_g} - 1$, $\Xi_2 =  \frac{P_{o}^{{\max}} - P_g \|{\bf{\Theta}}_r {\textbf{h}}_{go}\|^2 - P_j \|{\bf{\Theta}}_r \textbf{H}_{bo} \textbf{w}_t \|^2 -
\|{\bf{\Theta}}_t\|_{F}^2 \sigma_{o}^{2} - \|{\bf{\Theta}}_r\|_{F}^2 \sigma_{o}^{2}}{\|{\bf{\Theta}}_t {\textbf{h}}_{ao}\|^2}$, and $\Xi_3 =\frac{  (\kappa - 1)\left(\sigma_{w}^{2} + P_g  |{\textbf{g}}_{gw}|^2 + P_j |\textbf{g}_{bw} \textbf{w}_t|^2 + \|{\textbf{h}}_{ow} {\bf{\Theta}}_r\|^2 \sigma_{o}^{2} \right) - \|{\textbf{h}}_{ow} {\bf{\Theta}}_t\|^2 \sigma_{o}^{2}}{|{\textbf{g}}_{aw}|^2}$.
Due to the fact $\tfrac{\partial R_a}{\partial P_a} > 0$, the optimal transmission power of Alice can be given by
\begin{align}
P_{a}^{*} = {\min} \{P_{a}^{{\max}}, \Xi_1, \Xi_2, \Xi_3 \}. \label{P_b_O}
\end{align}
Similarly,  with any given $P_a$ and $P_g$, constraints \eqref{P1b}, \eqref{P1d}, \eqref{P_s}, and \eqref{PbPg_k} can be written as $P_j \le P_{j}^{{\max}}$, $P_j \le \Xi_4$, $P_j \le \Xi_5$, and $P_j \ge \Xi_6$, where $\Xi_4 = \frac{P_g|\textbf{g}_{gb}|^2 /\mu_g - P_a |\textbf{g}_{ab}|^2 - \|\textbf{w}_{r}^{H} \textbf{H}_{ob} {\bf{\Theta}}_t \|^2 \sigma_{o}^{2} - \|\textbf{w}_{r}^{H} \textbf{H}_{ob} {\bf{\Theta}}_r \|^2 \sigma_{o}^{2} - \sigma_{b}^{2}}{\phi |\textbf{w}_{r}^{H}\textbf{H}_{bb} \textbf{w}_t|^2 + \phi |\textbf{w}_{r}^{H} \textbf{H}_{ob} {\bf{\Theta}}_r \textbf{H}_{bo} \textbf{w}_t|^2}$, $\Xi_5 = \frac{P_{o}^{\max} - P_g \|{\bf{\Theta}}_r {\textbf{h}}_{go}\|^2 - P_a \|{\bf{\Theta}}_t {\textbf{h}}_{ao}\|^2 - \|{\bf{\Theta}}_t\|_{F}^2 \sigma_{o}^{2} - \|{\bf{\Theta}}_r\|_{F}^2 \sigma_{o}^{2} }{\|{\bf{\Theta}}_r \textbf{H}_{bo} \textbf{w}_t \|^2}$, and $\Xi_6 = \frac{(P_a |\textbf{g}_{aw}|^2 + \|\textbf{h}_{ow} {\bf{\Theta}}_t\|^2 \sigma_{o}^{2})/(\kappa - 1) - P_g|\textbf{g}_{gw}|^2 - \|\textbf{h}_{ow} {\bf{\Theta}}_r\|^2 \sigma_{o}^{2} - \sigma_{w}^{2}}{|\textbf{g}_{bw}\textbf{w}_t|^2}$. Due to the fact   $\frac{\partial R_a}{\partial P_j} < 0$, the optimal jamming power is given by

\begin{align}
P_{j}^{*} =
\left\{ {\begin{array}{*{20}{c}}
            {  \Xi_6
            }, &\Xi_6 \le {\rm{min}}\{P_{j}^{{\max}},\Xi_4,\Xi_5 \}, \\
            0, &{\rm{otherwise} }.
            \end{array}} \right. \label{P_j*}
\end{align}
In \eqref{P_j*}, when $\Xi_4 < \Xi_6$ or $\Xi_5 < \Xi_6$, the optimal jamming power does not exist. In this case, $P_j = 0 $ is set to ensure the covert communication performance. Furthermore, to confuse Willie's detection and satisfy Grace's QoS requirements, Grace's optimal transmit power is given by $P_{g}^{*} = P_{g}^{{\max}}$.

For any given transmit powers and Bob's FD beamforming, problem (P1) reduces to the sub-problem of optimizing the A-IOS's refraction and reflection beamforming. 
By introducing $\textbf{v}_t = [\alpha_{1}^{t} e^{j\theta_{1}^{t}}, \cdots, \alpha_{K}^{t} e^{j\theta_{K}^{t}}]^T$, $\textbf{v}_r = [\alpha_{1}^{r} e^{j\theta_{1}^{r}}, \cdots, \alpha_{K}^{r} e^{j\theta_{K}^{r}},1]^T$, $\textbf{V}_i = \textbf{v}_i \textbf{v}_{i}^H$, $i \in \{t,r \}$, $\textbf{A} = [\textbf{H}_{ob} {\rm{diag}}(\textbf{h}_{go}),\textbf{0}_{M \times 1}]$, $\textbf{B} = \textbf{H}_{ob} {\rm{diag}}(\textbf{h}_{ao})$, $\textbf{C} = [\textbf{h}_{go}^*;0]\odot [\textbf{h}_{ow}^{H};0]$, $\textbf{D} = \textbf{h}_{ao}^* \odot \textbf{h}_{ow}^{H}$, $\tilde{\textbf{E}} =[{\rm{diag}}(\textbf{h}_{ow}) \textbf{H}_{bo}\textbf{w}_t;\textbf{h}_{bw}\textbf{w}_t]$, and $\tilde{\textbf{F}} = [{\rm{diag}}(\textbf{w}_{r}^{H}\textbf{H}_{ob})\textbf{H}_{bo}\textbf{w}_t;\textbf{w}_{r}^{H} \textbf{H}_{bb} \textbf{w}_t]$, we construct $\textbf{W}_r = \textbf{w}_r \textbf{w}_{r}^{H}$, $\textbf{E} = \tilde{\textbf{E}}\tilde{\textbf{E}}^H$, and $\textbf{F} = \tilde{\textbf{F}}\tilde{\textbf{F}}^H$. 
Then, we have 
$H_{gb}  \triangleq  | \textbf{g}_{gb} |^2  
= {\rm{tr}}(\textbf{A}\textbf{V}_r\textbf{A}^H\textbf{W}_r) $, $ H_{ab} \triangleq  | \textbf{g}_{ab} |^2 = {\rm{tr}}(\textbf{B}\textbf{V}_t\textbf{B}^H\textbf{W}_r) $, $H_{gw} \triangleq |\textbf{g}_{gw} |^2 = {\rm{tr}}(\textbf{C}\textbf{C}^H \textbf{V}_r) $, $H_{aw} \triangleq  |\textbf{g}_{aw} |^2 = {\rm{tr}}(\textbf{D}\textbf{D}^H \textbf{V}_t) $, $H_{bw} \triangleq |\textbf{g}_{bw} \textbf{w}_t|^2 = {\rm{tr}}(\textbf{E}\textbf{V}_{r}) $, and ${H}_{bb} \triangleq |\textbf{w}_{r}^{H}(\textbf{H}_{bb} + \textbf{H}_{ob}{\bf{\Theta}}_r \textbf{H}_{bo}) \textbf{w}_t|^2 = {\rm{tr}}(\textbf{F}\textbf{V}_r ) $. Furthermore, by introducing ${\bf{\Phi}}_{rw} = {\rm{diag}}(|[{\textbf{h}}_{ow}]_1|^2,\cdots,|[\textbf{h}_{ow}]_K|^2,0)$, ${\bf{\Phi}}_{tw} = {\rm{diag}}(|[{\textbf{h}}_{ow}]_1|^2,\cdots,|[\textbf{h}_{ow}]_K|^2)$, ${\bf{\Phi}}_{go} =  {\rm{diag}}(|[{\textbf{h}}_{go}]_1|^2, $ $\cdots,|[\textbf{h}_{go}]_K|^2,0)$, 
${\bf{\Phi}}_{ao} =   {\rm{diag}}(|[{\textbf{h}}_{ao}]_1|^2,\cdots,|[\textbf{h}_{ao}]_K|^2)$, ${\bf{\Phi}}_{rb} = {\rm{diag}}(|[\textbf{w}_{r}^{H}{\textbf{H}}_{ob}]_1|^2,\cdots,|[\textbf{w}_{r}^{H}\textbf{H}_{ob}]_K|^2,0)$, ${\bf{\Phi}}_{tb} = $ $  {\rm{diag}}( |[\textbf{w}_{r}^{H}{\textbf{H}}_{ob}]_1|^2,\cdots,|[\textbf{w}_{r}^{H}\textbf{H}_{ob}]_K|^2)$,
${\bf{\Phi}}_{bo} = {\rm{diag}} $ $    ( |[{\textbf{H}}_{bo}\textbf{w}_{t}]_1|^2, \cdots,|[\textbf{H}_{bo}\textbf{w}_{t}]_K|^2,0)$, 
and ${\bf{\Pi}} = {\rm{diag}} $ $ \big([{\textbf{1}}_{1 \times K},0]^T \big)$,  
we have $\|\textbf{h}_{ow} {\bf{\Theta}}_i \|^2 = {\rm{tr}}({\bf{\Phi}}_{iw} \textbf{V}_i)$, $\|\textbf{w}_{r}^{H}\textbf{H}_{ob} {\bf{\Theta}}_i \|^2 = {\rm{tr}}({\bf{\Phi}}_{ib} \textbf{V}_i)$, $i \in \{t, r \}$, $\|{\bf{\Theta}}_r \|_{F}^{2} = {\rm{tr}}({\bf{\Pi}} \textbf{V}_r)$, $\|{\bf{\Theta}}_r \textbf{h}_{go} \|^2 = {\rm{tr}}({\bf{\Phi}}_{go} \textbf{V}_r)$, $\|{\bf{\Theta}}_t \textbf{h}_{ao} \|^2 = {\rm{tr}}({\bf{\Phi}}_{ao} \textbf{V}_t)$, and $\|{\bf{\Theta}}_r \textbf{H}_{bo}\textbf{w}_t \|^2 = {\rm{tr}}({\bf{\Phi}}_{bo} \textbf{V}_r)$.
Let $\Omega_1 = {\rm{tr}}({\bf{\Phi}}_{rb} {\textbf{V}}_r) \sigma_{o}^{2} + {\rm{tr}}({\bf{\Phi}}_{tb} {\textbf{V}}_t) \sigma_{o}^{2} + \phi P_j {H}_{bb}$ and $\gamma_a (\textbf{V}_t, \textbf{V}_r) = \frac{P_a{H_{ab}}}{\Omega_1 + \sigma_{b}^{2}}$, the sub-problem of optimizing the A-IOS refraction and reflection beamforming can be equivalently written as: 
\begin{subequations}
\begin{align}
({\rm{P3}}):~&\mathop{{\max}}\limits_{{\textbf{V}}_r, {\textbf{V}}_t, p_k, p_{k}^{{\rm{in}}}} \gamma_a (\textbf{V}_t, \textbf{V}_r)  \label{P3a}\\
{\rm{s.t.}}~&P_g H_{gb} 
\ge \mu_g \big( P_a H_{ab} + \Omega_1 +\sigma_{b}^{2}  \big),  \label{P3b} \\ 
~& P_g {\rm{tr}}({\bf{\Phi}}_{go} {\textbf{V}}_r) + P_a {\rm{tr}}({\bf{\Phi}}_{ao} {\textbf{V}}_t) + {\rm{tr}} ( {\textbf{V}}_t) \sigma_{o}^{2} \nonumber \\
~& + {\rm{tr}} ({\bf{\Pi}} {\textbf{V}}_r) \sigma_{o}^{2} + P_j {\rm{tr}}({\bf{\Phi}}_{bo} \textbf{V}_r) \le P_{o}^{{\max}}, \label{P3c}
\\ ~& P_a H_{aw}  + {\rm{tr}}({\bf{\Phi}}_{tw} {\textbf{V}}_t)\sigma_{o}^{2} \le (\kappa - 1)  \nonumber \\ ~&  \times \left(P_g H_{gw}  + P_j H_{bw} + {\rm{tr}}({\bf{\Phi}}_{rw} {\textbf{V}}_r) \sigma_{o}^{2} + \sigma_{w}^{2} \right), \label{P3d}
\\ 
~& H_{gb}   \ge H_{ab}, 
  \label{P3e}
\\ ~& \textbf{V}_{i,(k,k)} \le (\alpha_{k}^{{\max}})^2, ~\forall k \in K, ~i \in \{t,r \},  \label{P3f}
\\ ~& {\textbf{V}}_t \succeq 0, {\textbf{V}}_r \succeq 0, ~\eqref{P1i},\eqref{P1j}, 
\text{~and~} \textbf{V}_{r,(K+1,K+1)} = 1, \label{P3g}
\\ ~& {\rm{rank}}({\textbf{V}}_t) = {\rm{rank}}({\textbf{V}}_r) = 1,\label{P3h}
\\ & \! \left(\textbf{V}_{t, (k,k)} + \textbf{V}_{r,(k,k)} \right) \le \tfrac{p_k}{p_{k}^{{\rm{in}}} + \sigma_{o}^{2}}.
\label{P3i}
\end{align}
\end{subequations}
Now, problem (P3) is a concave-convex fractional programming problem. To tackle problem (P3), we employ the Dinkelbach transform by first converting $\gamma_a (\textbf{V}_t, \textbf{V}_r)$ to
\begin{align}
\widehat{\gamma}_{a}^{m+1} (\textbf{V}_t, \textbf{V}_r) = ~&  {P_a}{\rm{tr}}(\textbf{B} \textbf{V}_t \textbf{B}^H \textbf{W}_r) - \eta_{1}^{m}\big({\rm{tr}}({\bf{\Phi}}_{tb} \textbf{V}_t)\sigma_{o}^{2}  \nonumber \\ & + {\rm{tr}}({\bf{\Phi}}_{rb}\textbf{V}_{r} )\sigma_{o}^{2} + \phi P_j{\rm{tr}}(\textbf{F}\textbf{V}_r) + \sigma_{b}^{2} \big), \label{gamma_b}   
\end{align}
where $m$ is the iteration index and the control factor $\eta_{1}^{m}$ is updated by
\begin{align}
\eta_{1}^{m} = ~&  {P_a}{\rm{tr}}(\textbf{B} \textbf{V}_{t}^{m} \textbf{B}^H \textbf{W}_r) \big( {\rm{tr}}({\bf{\Phi}}_{tb} \textbf{V}_{t}^{m})\sigma_{o}^{2} + {\rm{tr}}({\bf{\Phi}}_{rb}\textbf{V}_{r}^{m} )\sigma_{o}^{2}  \nonumber \\ & +  \phi P_j{\rm{tr}}(\textbf{F}\textbf{V}_r) + \sigma_{b}^{2} \big)^{-1}. 
\label{eta_m}
\end{align}
$\!$In addition, $\eta_{1}^{m}$ is updated until it stops at $\widehat{\gamma}_{a}^{m+1} (\textbf{V}_t, \textbf{V}_r)\le 0$.  
To deal with the non-convex constraint \eqref{P3i}, we apply the arithmetic-geometric mean inequality to approximate it as:
\begin{align} 
\big(\big( \textbf{V}_{t,(k,k)} + \textbf{V}_{r,(k,k)}\big)\mu_1 \big)^2 + \big((p_{k}^{{\rm{in}}} + \sigma_{o}^{2})/\mu_1 \big)^2 \le 2p_k, \label{AGM}
\end{align}
where $\mu_1 = \sqrt{(p_{k}^{{\rm{in}}} + \sigma_{o}^{2})/(\textbf{V}_{t,(k,k)} + \textbf{V}_{r,(k,k)})}$.
Next, we deal with the rank-one constraint \eqref{P3h} by using the following fact:
\begin{align}
{\rm{rank}}(\textbf{V}_i) = 1 \Longleftrightarrow \varsigma_i \triangleq {\rm{tr}}(\textbf{V}_i) - \|\textbf{V}_i \|_2 = 0, i\in \{t, r \},      
\end{align}
where $\|\cdot\|_2$ stands for the spectral norm. 
Note that when ${\rm{rank}} ({\textbf{V}}_i) = 1$, ${\rm{tr}}(\textbf{V}_i) - \|\textbf{V}_i \|_2 = 0 $ always holds true; Otherwise, ${\rm{tr}}(\textbf{V}_i) - \|\textbf{V}_i \|_2 \ge 0 $ always holds true for any positive semidefinite matrix $\textbf{V}_i$. Thus, we add the rank-one constraint as a penalty term into the objective function and arrive at the following optimization problem:
\begin{subequations}
\begin{align}
({\rm{P4}}):~&\mathop{{\max}}\limits_{{\textbf{V}}_r, {\textbf{V}}_t} ~ \widehat{\gamma}_{a}^{m+1}(\textbf{V}_t, \textbf{V}_r) - \tfrac{1}{\rho_1} (\varsigma_t + \varsigma_r)
 \label{P5a}\\
{\rm{s.t.}}~& \eqref{P3b}, \eqref{P3c}, \eqref{P3d},  \eqref{P3e}, \eqref{P3f}, \eqref{P3g}, \text{~and~} \eqref{AGM}.
\end{align}
\end{subequations}
In the above objective function, $\rho_1 \ge 0$ is the penalty control factor. However, the objective function of problem (P4) is still non-concave due to the term $\|{\textbf{V}}_i\|_2$ involved in $\varsigma_i$. Then, we apply the first-order Taylor expansion of $\varsigma_i$ to obtain its upper bound as:
\begin{align} 
\varsigma_i \le &~   {\rm{tr}}(\textbf{V}_i)  - \left(\|\textbf{V}_{i}^{n}\|_2 + {\rm{tr}} \left( \textbf{v}_{{\max},i}^{n} (\textbf{v}_{{\max},i}^{n})^H (\textbf{V}_{i} -\textbf{V}_{i}^{n}) \right)  \right) \nonumber \\ 
\triangleq &~ \hat{\varsigma}_{i}(\textbf{V}_i),~ i \in \{t,r \}, \label{up}
\end{align}
where $\textbf{v}_{{\max},i}^{n}$ is the eigenvector corresponding to the largest eigenvalue of $\textbf{V}_{i}^{n}$ and $n$ is the number of the iterations. By substituting $\hat{\varsigma}_{i}(\textbf{V}_i)$ into the objective function of problem (P4), we formulate the following optimization problem as:
\begin{subequations}
\begin{align}
({\rm{P5}}):~&\mathop{{\max}}\limits_{{\textbf{V}}_r, {\textbf{V}}_t} ~ \widehat{\gamma}_{a}^{m+1}(\textbf{V}_t, \textbf{V}_r) - \tfrac{1}{\rho_1} (\hat{\varsigma}_{t}(\textbf{V}_t) + \hat{\varsigma}_{r}(\textbf{V}_r))
 \label{P5a}\\
{\rm{s.t.}}~& \eqref{P3b},\eqref{P3c},\eqref{P3d}, \eqref{P3e},\eqref{P3f},\eqref{P3g}, \text{~and~} \eqref{AGM}.
\end{align}
\end{subequations}
Now, problem (P5) is a semidefinite programming, which can be solved by a convex optimization solver, such as CVX. Note that the penalty control factor in problem (P5) is update by $\rho_1 = c_1 \rho_1$  with  $0< c_1 <1$ untill $\frac{1}{\rho_1} (\hat{\varsigma}_t   + \hat{\varsigma}_r  ) \le \zeta_1$. 

For any given A-IOS beamforming, NOMA transmit power, and FD jamming power, problem (P1) can be reduced to optimize Bob's FD beamforming. To this end, we introduce $\textbf{G}_b = ((\textbf{H}_{bb}+\textbf{H}_{ob}{\bf{\Theta}}_r \textbf{H}_{bo})\textbf{w}_t)((\textbf{H}_{bb}+\textbf{H}_{ob}{\bf{\Theta}}_r \textbf{H}_{bo})\textbf{w}_t)^H$ and ${\textbf{G}}_i = ({\textbf{H}}_{ob} {\bf{\Theta}}_i) ({\textbf{H}}_{ob} {\bf{\Theta}}_i)^H, i\in \{t,r \}$ and obtain  $|\textbf{w}_{r}^{H} (\textbf{H}_{bb}+\textbf{H}_{ob}{\bf{\Theta}}_r \textbf{H}_{bo})\textbf{w}_t|^2 = {\rm{tr}}(\textbf{G}_b \textbf{W}_r)$ and $\|{\textbf{w}}_{r}^{H} {\textbf{H}}_{ob} {\bf{\Theta}}_i\|^2 = {\rm{tr}}({\textbf{G}}_i {\textbf{W}_r})$. 
Then, let $\Omega_2 = {\rm{tr}}( {\textbf{G}}_r {\textbf{W}_r}) \sigma_{o}^{2} +  {\rm{tr}}( {\textbf{G}}_t {\textbf{W}_r}) \sigma_{o}^{2}$, the sub-problem of optimizing Bob's receive beamforming can be written as: 
\begin{subequations}
\begin{align}
({\rm{P6}}): ~&\mathop{{\max}}\limits_{{\textbf{W}_r}} ~\tfrac{{P_a}{\rm{tr}}(\textbf{B}\textbf{V}_t\textbf{B}^H {\textbf{W}_r})}{ \Omega_2 + \phi P_j {\rm{tr}}(\textbf{G}_b \textbf{W}_r)  + \sigma_{b}^{2}} \label{P6a}\\
{\rm{s.t.}} ~& P_g {\rm{tr}}(\textbf{A}\textbf{V}_r\textbf{A}^H {\textbf{W}_r}) \ge \mu_g \big(P_a {\rm{tr}}(\textbf{B}\textbf{V}_t\textbf{B}^H {\textbf{W}_r}) + \Omega_2  \nonumber \\
~&  + \phi P_j{\rm{tr}}(\textbf{G}_b \textbf{W}_r) + \sigma_{b}^{2} \big), \label{P6b}  \\~&{\rm{tr}}({\textbf{W}_r}) = 1 \text{~and~} {\textbf{W}_r} \succeq 0,    \label{P6c} \\~&{\rm{rank}}({\textbf{W}_r}) = 1. \label{P6d}
\end{align}
\end{subequations}
\begin{algorithm}[t]
\footnotesize
 \caption{Proposed AO Algorithm}
 \label{alg1}
 \begin{algorithmic}[1]
  \STATE Initialize the feasible $\ell \gets 0$, $P_{a}^{0}$, $P_{g}^{0}$, ${\textbf{V}}_{t}^{0}$, ${\textbf{V}}_{r}^{0}$, $\textbf{W}_{t}^{0}$, and ${\textbf{W}}_{r}^{0}$.
  \REPEAT 
  \STATE Update $P_{a}^{\ell+1} = {\rm{min}} \{P_{a}^{{\max}},\Xi_2,\Xi_3,\Xi_4 \}$, $ P_{g}^{\ell+1} = P_{g}^{\max}$, $n \gets 0$.
  \REPEAT
  \STATE Initialize $\eta_{1}^{0} > 0, m \gets 0$.
  \REPEAT 
  \STATE Update ${\textbf{V}}_{r}^{m+1}$ and ${\textbf{V}}_{t}^{m+1}$ by solving (P5).
  \STATE Update $\eta_{1}^{m+1}$ by \eqref{eta_m}. $m = m+1$.
  \UNTIL $\widehat{\gamma}_{a}^{m+1}(\textbf{V}_t, \textbf{V}_r) \le 0$.
  \STATE ${\textbf{V}}_{r}^{n} \gets {\textbf{V}}_{r}^{m}$, ${\textbf{V}}_{t}^{n} \gets {\textbf{V}}_{t}^{m}$. $\rho_1 \gets c_1 \rho_1$, $n = n+1$.
  \UNTIL $\tfrac{1}{\rho_1} (\hat{\varsigma}_t   + \hat{\varsigma}_r  ) \le \zeta_1$.
  \STATE ${\textbf{V}}_{r}^{\ell} \gets {\textbf{V}}_{r}^{n}$, ${\textbf{V}}_{t}^{\ell} \gets {\textbf{V}}_{t}^{n}$.  $n \gets 0$.
  \REPEAT
  \STATE Initialize $\eta_{2}^{0} > 0, m \gets 0$.
  \REPEAT
  \STATE Update $\textbf{W}_r^{m+1}$ by solving (P7).
  \STATE Update $\eta_{2}^{m} = f_{1}^{m} (\textbf{W}_r) / f_{2}^{m} (\textbf{W}_r)$. $m = m + 1$.
  \UNTIL $f_{1}^{m} (\textbf{W}_r) - \eta_{2}^{m} f_{2}^{m} (\textbf{W}_r) \le 0$.
  \STATE ${\textbf{W}_r}^{n} \gets {\textbf{W}_r}^{m}$. $\rho_2 \gets c_2 \rho_2$. $n = n+1$.
  \UNTIL $\tfrac{1}{\rho_2} ({\rm{tr}}(\textbf{W}_{r}^{n})  + \widehat{\textbf{W}}_{r}^{n} ) \le \zeta_2$.
  \STATE ${\textbf{W}}_{r}^{\ell} \gets {\textbf{W}}_{r}^{n}$, $n \gets 0$.
  \REPEAT 
  \STATE Update $\textbf{W}_{t}^{n}$ by solving (P8), $\rho_3 \gets c_3\rho_3$.
  \UNTIL $\frac{1}{\rho_3} ({\rm{tr}}(\textbf{W}_{t}^{n})  + \widehat{\textbf{W}}_{t}^{n} ) \le \zeta_3$.
  \STATE ${\textbf{W}}_{t}^{\ell} \gets {\textbf{W}}_{t}^{n}$, $\ell = \ell + 1$.
  \UNTIL $R_{a}^{\ell+1} - R_{a}^{\ell} < \zeta_4$.
 \end{algorithmic}
\end{algorithm}
Similarly to the procedures of solving problems (P4) and (P5), we introduce $\widehat{{\textbf{W}}}_i^{n} \triangleq -\|\textbf{W}_i^{n}\|_2 - {\rm{tr}} \left( \textbf{w}_{i,{\max}}^{n} (\textbf{w}_{i,{\max}}^{n})^H (\textbf{W}_i -\textbf{W}_i^{n}) \right), i \in \{t,r \}$ and define $f_1(\textbf{W}_r)  = {P_a}{\rm{tr}}(\textbf{B}\textbf{V}_t \textbf{B}^H {\textbf{W}_r})$ and $f_2(\textbf{W}_r) = \Omega_2 $ $ + \phi P_j {\rm{tr}}(\textbf{G}_b \textbf{W}_r)  + \frac{1}{\rho_2} 
({\rm{tr}}(\textbf{W}_r) + \widehat{\textbf{W}}_r^{n} ) + \sigma_{b}^{2} $. Then, the sub-problem of optimizing Bob's receive beamforming can be expressed as: 
\begin{subequations}
\begin{align}
({\rm{P7}}):~&\mathop{{\max}}\limits_{{\textbf{W}_r}} ~f_{1}(\textbf{W}_r) - \eta_{2}^{m} f_2(\textbf{W}_r) \label{P7a}\\
{\rm{s.t.}}~& \eqref{P6b} 
\text{~and~} \eqref{P6c}.
\end{align}
\end{subequations} 
As such, we introduce $\textbf{W}_t = \textbf{w}_t \textbf{w}_{t}^{H}$, $\tilde{\textbf{G}}_r = ({\bf{\Theta}}_r \textbf{H}_{bo})^H {\bf{\Theta}}_r \textbf{H}_{bo}$, $\textbf{G}_{j} = \textbf{w}_{r}^{H}(\textbf{H}_{bb} + \textbf{H}_{ob} {\bf{\Theta}}_r \textbf{H}_{bo})$, and $\tilde{\textbf{G}}_w = \textbf{g}_{bw}^{H}\textbf{g}_{bw}$ and obtain $\|{\bf{\Theta}}_r \textbf{H}_{bo} \textbf{w}_t \|^2 = {\rm{tr}}(\tilde{\textbf{G}}_r \textbf{W}_t)$, $|\textbf{g}_{bw}\textbf{w}_t|^2 = {\rm{tr}}(\tilde{\textbf{G}}_w \textbf{W}_t)$, $\tilde{\textbf{G}}_j = \textbf{G}_{j}^{H}\textbf{G}_j$, and $|\textbf{w}_{r}^{H}(\textbf{H}_{bb}+ \textbf{H}_{ob}{\bf{\Theta}}_r\textbf{H}_{bo}) \textbf{w}_t|^2 = {{\rm{tr}}}( \tilde{\textbf{G}}_{j} \textbf{W}_t)$. By transforming the constraint ${\rm{rank}}(\textbf{W}_t) = 1$ into the objective function, $|\textbf{w}_{r}^{H} (\textbf{H}_{bb} + \textbf{H}_{ob}{\bf{\Theta}}_r \textbf{H}_{bo} ) \textbf{w}_t|^2$, as a penalty term, the sub-problem of optimizing Bob's transmit beamforming can be written as: 
\begin{subequations}
\begin{align}
({\rm{P8}}):~&\mathop{{\rm{min}}}\limits_{{\textbf{W}_t}} ~{\rm{tr}}(\tilde{\textbf{G}}_j \textbf{W}_t) + \tfrac{1}{\rho_3}({\rm{tr}}(\textbf{W}_t) + \widehat{\textbf{W}}_{t}^{n}) \label{P8a}\\
{\rm{s.t.}}~& P_a H_{aw}  + {\rm{tr}}({\bf{\Phi}}_{tw} {\textbf{V}}_t)\sigma_{o}^{2} \le (\kappa - 1) \times \nonumber    \big(P_g H_{gw}\\ ~&  + P_j {\rm{tr}}(\tilde{\textbf{G}}_w \textbf{W}_t) + {\rm{tr}}({\bf{\Phi}}_{rw} {\textbf{V}}_r) \sigma_{o}^{2} + \sigma_{w}^{2} \big),    \label{P8b}  
\\~&P_g {\rm{tr}}({\bf{\Phi}}_{go} {\textbf{V}}_r) + P_a {\rm{tr}}({\bf{\Phi}}_{ao} {\textbf{V}}_t) + {\rm{tr}} ({\textbf{V}}_t) \sigma_{o}^{2} \nonumber \\ 
~& + {\rm{tr}} ({\bf{\Pi}} {\textbf{V}}_r) \sigma_{o}^{2} + P_j {\rm{tr}}(\tilde{\textbf{G}}_r \textbf{W}_t) \le P_{o}^{{\max}},\label{P8c}
\\~&{\rm{tr}}({\textbf{W}_t}) = 1 \text{~and~} {\textbf{W}_t} \succeq 0.\label{P8d}
\end{align}
\end{subequations}
Now, problems (P7) and (P8) are semidefinite programming, which can be solved by using the standard convex solvers. 
 
The proposed AO algorithm is summarized Algorithm 1 and the corresponding complexity mainly comes from solving the semidefinite programming subproblems (P5), (P7), and (P8), which contain the corresponding penalty term iterations. Thus, the complexity of solving the problems (P5), (P7), and (P8) are $\mathcal{O} \left(I_1 I_2 I_3 \left(K^{3.5} \right)\right)$ and $\mathcal{O} \left(I_1 I_2 I_3 \left(M^{3.5} \right)\right)$, and  $\mathcal{O} \left(I_1 I_2 \left(M^{3.5} \right)\right)$, respectively, where $I_1$, $I_2$, and $I_3$ are the numbers of the AO iterations, penalty term iterations, and Dinkelbach transform iterations, respectively.

\section{Simulation Results}
  
In simulations, unless otherwise stated, the simulation parameters are set as $M = 4$, $K = 16$, $\sigma_{b}^{2} = \sigma_{o}^{2} = \sigma_{w}^{2} = -90$ dBm, $\hat R_{g} = 1$ bps/Hz, $\zeta_1= \zeta_2 = \zeta_3 = 10^{-4}$, $\zeta_4 = 10^{-2}$ \cite{Zhumiaomiao}, $P_{a}^{{\max}} = P_{g}^{{\max}}$, $P_{o}^{{\max}} = 20$ dBm, and the SI cancellation level is $\phi = -140$ dB \cite{STAR_RIS_FD_RSMA}. 
Consider a two-dimensional coordinate area, 
Alice, Willie, A-IOS, Bob, and Grace are located at (70 m, 15 m), (20 m, -5 m), (35 m, 5 m), (0 m, 0 m), and (70 m, -5 m), respectively. 
The channel fading coefficients are modeled as the same as that in \cite{Zhumiaomiao}. In all the figures, “A” and “P” denote the A-IOS and passive-IOS (P-IOS), respectively. For the comparison purpose, we consider the following benchmark schemes: 1) The half-duplex (HD) receiver scheme in which the  receiver does not conduct the FD jamming, i.e., $P_{j}^{{\max}} = 0$ W. 2) The A-IOS scheme with random ${\bf{\Theta}}_i$, $i \in \{t, r\}$ . 3) The P-IOS scheme in which the transmit power budget of Grace is $\tilde{P}_{g}^{{\max}} = {P}_{g}^{{\max}} + P_{o}^{{\max}}$ for a fair comparison.

In Fig. 1, the impacts of the jamming power budget on the covert rate are investigated. The curves in Fig. 1 show that the proposed AO algorithm with the A-IOS setting achieves a higher covert rate than the P-IOS setting. Compared to the random  ${\bf{\Theta}}_i$ scheme, the effectiveness of the A-IOS beamforming optimization is also verified. Moreover, with the increase of $P_j^{\max}$, the covert rates achieved by the proposed AO algorithm increase at first and then approach a constant in the high $P_j^{\max}$ region, which highlights that a middle value of $P_j^{\max}$ is enough to obtain the highest $R_a$. Compared to the HD receiver scheme ($P_{j}^{{\max}} = 0$ W), the significant improvements on $R_a$ can be achieved by the proposed FD receiver scheme. 

\begin{figure}[t]
	\centering
	\includegraphics[width=3in]{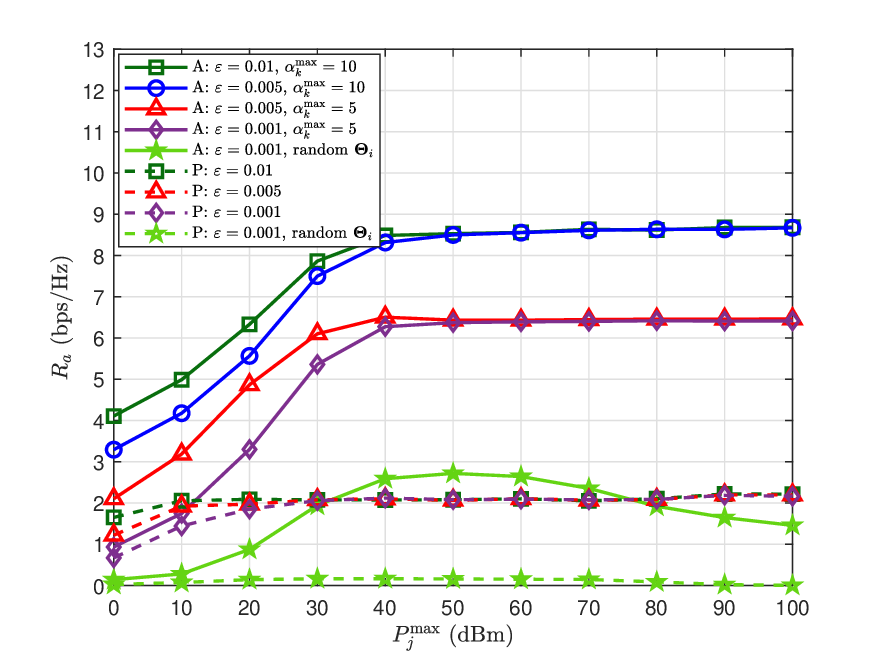}
        \vspace{-0.1in}
	\caption{Covert rate versus $P_{j}^{\max}$.}
	\label{Fig1}
\vspace{-0.15in}
\end{figure}
 
Fig. 2(a) and Fig. 2(b) depict $R_a$ versus $P_a^{\max}$ and $\alpha_{k}^{\max}$, respectively. 
In Fig. 2(a), all the curves of $R_a$ increase with the increasing $P_a^{\max}$. The A-IOS schemes achieve the higher $R_a$ than the corresponding P-IOS schemes. Also, the FD receiver with jamming obtains a higher $R_a$ scheme. The curves in Fig. 2(b) show that a relatively larger amplification amplitude $\alpha_k$ obtains a higher $R_a$ for the A-IOS schemes. Moreover, $R_a$ can be effectively improved by increasing the number of the A-IOS elements. 

\begin{figure}[t]
	\centering
	\includegraphics[width=3in]{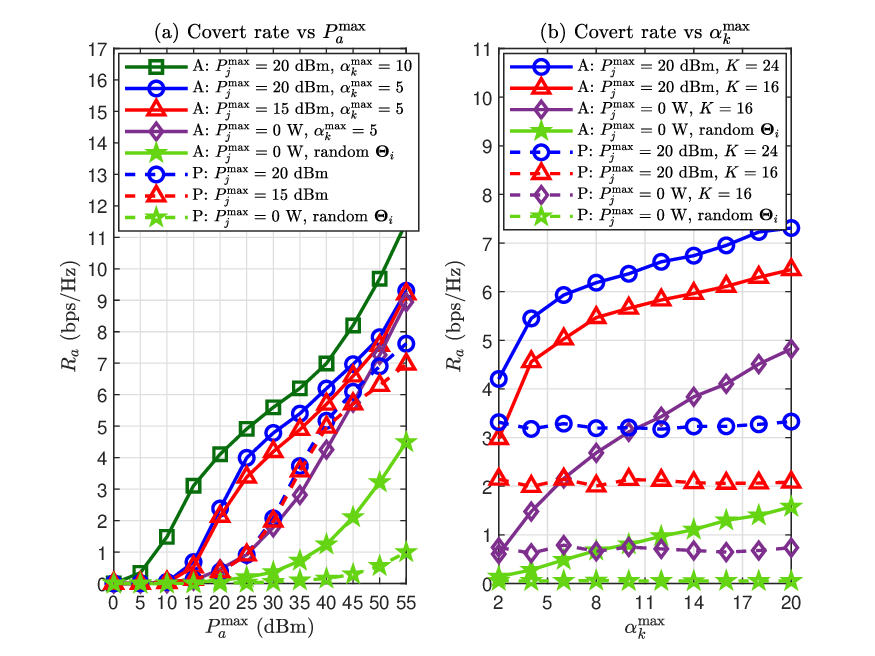}
        \vspace{-0.1in}
	\caption{Covert rate versus $P_{a}^{\max}$ and $\alpha_{k}^{{\max}}$.}
	\label{Fig2}
\vspace{-0.15in}
\end{figure}

\begin{figure}[t]
	\centering
	\includegraphics[width=3in]{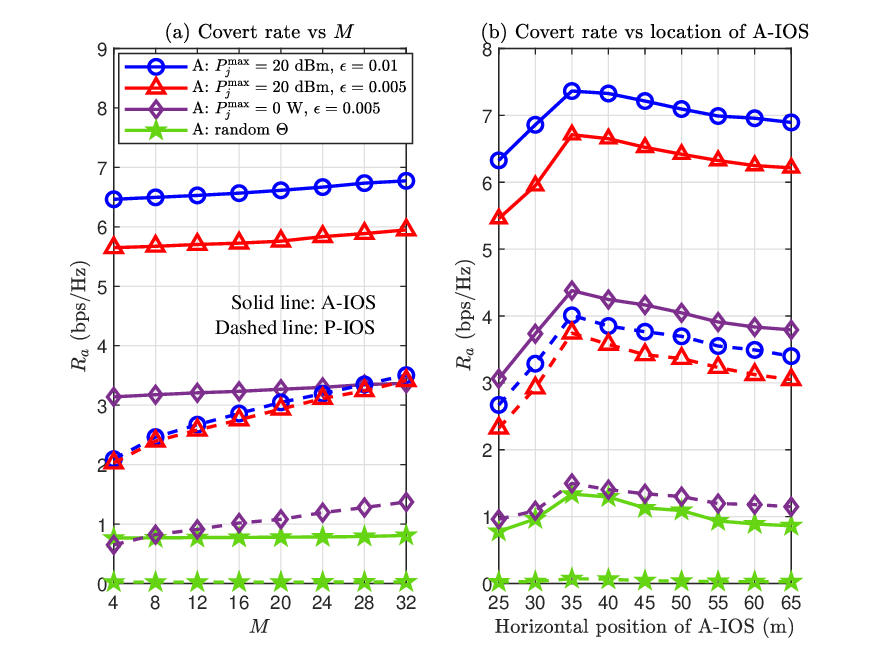}
        \vspace{-0.1in}
	\caption{Covert rate versus $M$ and location of A-IOS.}
	\label{Fig3}
\vspace{-0.15in}
\end{figure}

Figs. 3(a) and 3(b) depict the covert rate versus the number of the transmit/receive antennas and location of the A-IOS, respectively. 
In Fig. 3(a), all the curves of $R_a$ increase with the increasing $M$.
Moreover, the curves in Fig. 3(a) show that the covert rate of the P-IOS schemes with $P_j^{\max} = 20$ dBm is no less than that of the A-IOS scheme with $P_j^{\max} = 0$ W when $M \ge 28$.
Also, Fig. 3(a) shows that $R_a$ is significantly improved by employing the FD receiver with jamming.
The curves in Fig. 3(b) show that the highest $R_a$ is achieved by the proposed scheme when A-IOS is properly located close to Bob and Willie.
Moreover, $R_a$ achieved by the P-IOS scheme is far less than that achieved by the A-IOS scheme.

\section{Conclusions}

In this paper, we have proposed an A-IOS and FD receiver aided scheme to enhance the covert communication performance for the uplink NOMA system. An AO algorithm has been designed to obtain the optimal FD beamforming, A-IOS beamforming, NOMA transmit power, and FD jamming power. Simulation results have demonstrated the effectiveness of the proposed AO algorithm. It has been shown that the covert communication performance can be significantly improved by employing not only the A-IOS, but also the FD receiver.

\begin{balance}
\bibliography{IEEEabrv,IEEE_bib}
\end{balance}

\end{document}